1# The Integrity of Machine Learning Algorithms against Software Defect Prediction

Param Khakhar and, Rahul Kumar Dubey, *Senior Member IEEE***Abstract—** The increased computerization in recent years has resulted in the production of a variety of different software, however measures need to be taken to ensure that the produced software isn't defective. Many researchers have worked in this area and have developed different Machine Learning-based approaches that predict whether the software is defective or not. This issue can't be resolved simply by using different conventional classifiers because the dataset is highly imbalanced i.e the number of defective samples detected is extremely less as compared to the number of non-defective samples. Therefore, to address this issue, certain sophisticated methods are required. The different methods developed by the researchers can be broadly classified into Resampling based methods, Cost-sensitive learning-based methods, and Ensemble Learning. Among these methods. This report analyses the performance of the Online Sequential Extreme Learning Machine (OS-ELM) proposed by Liang *et.al.* against several classifiers such as Logistic Regression, Support Vector Machine, Random Forest, and Naïve Bayes after oversampling the data. OS-ELM trains faster than conventional deep neural networks and it always converges to the globally optimal solution. A comparison is performed on the original dataset as well as the over-sampled data set. The oversampling technique used is Cluster-based Over-Sampling with Noise Filtering. This technique is better than several state-of-the-art techniques for oversampling. The analysis is carried out on 3 projects KC1, PC4 and PC3 carried out by the NASA group. The metrics used for measurement are *recall* and *balanced accuracy*. The results are higher for OS-ELM as compared to other classifiers in both scenarios.

**Index Terms**— Class Imbalanced Problem, Online Sequential Extreme Learning Machine, Oversampling, Software Defect Prediction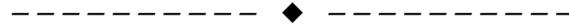

## 1 INTRODUCTION

CLASSIFIERS achieve good accuracy in predicting the unknown class if the training dataset is balanced however, in many scenarios this isn't the case. Software Defect Prediction is one such area where the number of non-defective instances is very high as compared to the number of defective instances. Thus, there is a skew in the classes. Machine Learning algorithms can be used to predict whether a given software is defective or non-defective based on the values of various features. However, when trained on skewed datasets they may be biased towards non-defective instances and fail to recognize defective instances. This is the main problem with using machine learning algorithms on such datasets and consequently, many researchers have worked for improving the performance of algorithms. The methods developed by the researcher to solve this problem can be classified into three main categories, Sampling-based methods, Cost-sensitive methods, and methods involving ensemble learning. Sampling-based methods involve undersampling from the majority class or over-sampling from the minority class. Cost-sensitive methods have different misclassification costs for the classifiers and thus penalize the classifiers non-uniformly while training. Ensemble learning involves aggregating the results of more than one learning algorithm.

Using deep neural network architectures for classification on software data may take a very long time for training and there is a high chance of overfitting due to the increased complexity of the model. OS-ELM developed by Liang *et.al* [1] addresses this issue. It is a simple learning algorithm for Single-Layer Feed-Forward Neural Network (SLFN). In theory, it provides good performances at a very fast learning speed. Moreover, it also provides better predictions for out of range instances thus increasing the reliability of the learning algorithm.

The oversampling technique used is KMFOS proposed by Gong *et.al* [2]. There are many oversampling based approaches developed by researchers. These techniques need to generate new minority instances systematically. The generation of the synthetic minority instances should neither be too random so that the classifier may fail to classify the majority class instances nor it should be too constrained near the existing minority instances so that the classifier overfits the dataset. The methods developed generally fail to completely achieve both the conditions. Additionally, the oversampling methods need to ensure that the generated instances don't result in an ambiguous decision boundary i.e. generation of a minority sample where the majority of the nearest neighbors are from the majority class. This would also affect the classifier performance. The KMFOS method tries to resolve the above-mentioned challenges in some way. The method comprises the following steps:

- The new instances generated by KMFOS are interpolation of the instances in the two clusters and thus the generated instances are distributed diversely.

- CLNI step removes potential outliers and thus helps in enhancing the performance of the classifier.

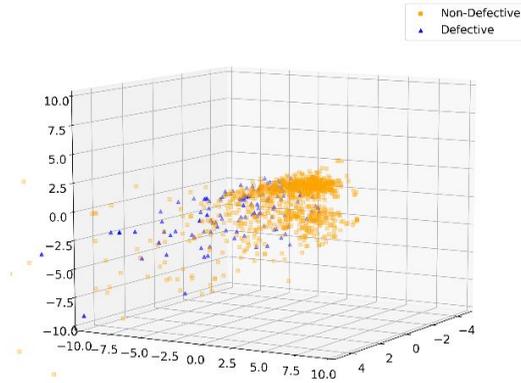

**Fig.1:** The dataset PC3 before over-sampling

Therefore, this technique is better than many state-of-the-art oversampling techniques. This report analyses the performance of the combination of an oversampling technique KMFOS and OS-ELM for software defect prediction as compared to other classifiers.

## 2 BACKGROUND

Researchers have used many sophisticated Machine Learning techniques for predicting whether the software is defective or not but, due to the nature of the datasets, the techniques fail to generalize and don't give optimal results. Over-sampling techniques result in the artificial creation of new minority instances to balance minority instances. Methods such as Bootstrapping which involve repeated sampling and introduction of the same samples lead to overfitting. Synthetic Minority Oversampling Technique introduced (SMOTE) proposed by Chawla *et.al* [3] uses Random Over Sampling as the core idea and generates new minority samples. This method fails to solve the ambiguous decision boundary problem. There are improved extensions to SMOTE such as Borderline SMOTE proposed by Han *et.al* [4], and SMOTE + ENN [5]. The method Adaptive Synthetic Sampling Approach for Imbalanced Learning was proposed by He *et.al* [6] generates synthetic instances proportional to the density of the minority instances in a region.

Under-sampling techniques such as random under-sampling, randomly select instances from the majority class. NearMiss proposed by Zhang *et.al* [7] uses the distance between the majority class and minority class examples to select instances that need to be discarded. Condensed Nearest Neighbor Rule proposed by Hart *et.al* [8] finds a minimal subset of the labels which results in no loss in model performance. Cost-sensitive training can be incorporated in the classifiers such as SVM, Logistic Regression, Decision Tree, etc. Cost-sensitive multilayer perceptron (CSMLP) is another learning algorithm which trains non-uniformly with different misclassification cost. Ensemble learning-based methods include aggregating several base classifiers using Bagging, Boosting, Gradient Boosting, XGBoost, etc.

## 3 PROPOSED METHOD
### 3.1 Overview

Firstly, results are computed using different classifiers using 5-Fold stratified cross-validation. Then the data is split in 5 folds followed by using KMFOS oversampling on each of the fold. The results are computed for each of the folds for each parameter of the KMFOS algorithm and averaged. Following classifiers were used in the study:

a) **Logistic Regression**

Logistic Regression is a discriminative classifier which tries to find a decision boundary in the feature space. The algorithm learns weights corresponding to each attribute. Classification is done based on the relative location of the instance concerning the decision boundary.

b) **Support Vector Machine**

For a feature space, more than 1 decision boundaries can exist. Support Vector Machines try to find a decision boundary that has the maximum margin i.e. The distance between the closest points from the decision boundary is maximized for points of both the class. These points are known as the support vectors and are used for making predictions for the new instances. If required, SVM can transform data into a higher dimensional space (using different kernel functions) to make it separable. Different kernel functions available in sklearn are linear, poly, sigmoid, and RBF. The values reported are the maximum for each kernel.

c) **Random Forest**

Random Forest is an ensemble of different Decision Trees. Each tree is trained on the bootstrapped sample of the original dataset. While constructing a split only a certain subset of all the features is considered. Increasing the number of features to be considered while making a split increases the strength of the individual trees. Increasing the height of the individual trees would also increase predictive strength. The number of trees used is crucial and depends on the strength of the individual trees. Therefore, there needs to be careful calibration of the hyperparameters to prevent overfitting. For uniformity, the parameters used are the default as available in the sklearn library.

d) **Naïve Bayes Classifier (Gaussian)**

Naïve Bayer Classifier is a generative classifier, which learns a Gaussian distribution and updates the values of the parameters with the data. The updation procedure is by the Bayes' Rule, wherein the prior is the original parameters of the distribution and the likelihood corresponds to the data which is received. The posterior again is a Gaussian distribution with the parameters changed.

**e) Online Sequential Extreme Learning Machine**

Online Sequential Extreme Learning Machine is a simple learning algorithm for Single-Layer Feed-Forward Neural Network (SLFN). Unlike traditional neural networks, OS-ELM does not use a gradient-based technique i.e. backpropagation for tuning the parameters on the contrary it tunes the parameters only once. The training of OS-ELM is carried out in two phases.

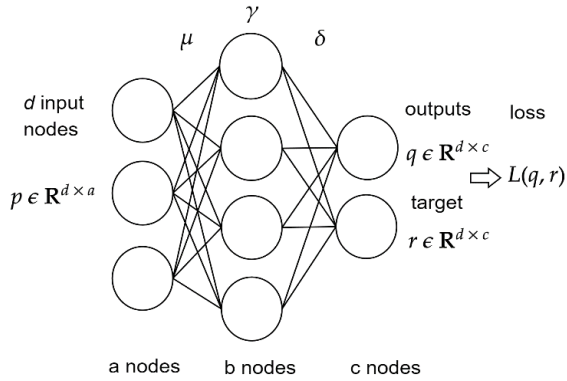

**Fig.2:** Network

1. Initial Training Phase

$$\mu, \delta, \gamma \Leftarrow \text{random input}$$
$$G_0 \Leftarrow H(p_0 \cdot \mu + \gamma) \quad d \geqslant b$$
$$K_0 \Leftarrow (G_0^T G_0)^{-1}$$
$$\delta_0 \Leftarrow K_0 G_0^T r_0$$

2. Sequential Training Phase

```
i ⇐ 0
for until (p_{i+1}, r_{i+1}) exists do
    K_{i+1} ⇐ K_i − K_i G_{i+1}^T (I + G_{i+1} K_i G_{i+1}^T)^{-1} G_{i+1} K_i
    δ_{i+1} ⇐ δ_i + K_{i+1} G_{i+1}^T (r_{i+1} − G_{i+1} δ_i)
    i ⇐ i + 1
end for
```

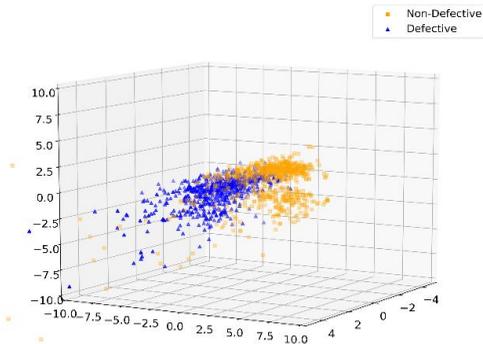

**Fig.3:** The dataset PC3 after over-sampling

In OS-ELM, the activation can only be used on the hidden nodes. OS-ELM is a special case of ELM. If all the samples are feed in the initial training phase of the network, it effectively becomes an ELM. Unlike other backpropagation based networks, OS-ELM does not train iteratively on the same data samples. Calculation of gradients is no longer required as the matrices are trained by computing some matrix products and matrix inversion. This results in the faster computation of the results. The algorithm involves matrix inversion whose complexity is O($N^3$), thus there is a constraint on the size of the batch which can be used. OS-ELM also provides better results for the instances which are out of range which is one of the requirements in software defect prediction.

### 3.2 KMFOS

KMFOS consists of three steps: clustering, over-sampling, and noise-filtering. First, K-means clustering is done on the entire dataset and as a result, the defective instances are also placed in the different clusters. Then all pairs of the clusters are considered and the minority instances in each of the clusters of a pair are taken for interpolating a new instance between them. In the filtering step, instances are removed whose majority of the nearest neighbors are of a different class than that of the instance.

### 3.2.1 Clustering

K-means is an unsupervised clustering algorithm where the data points are grouped into $k$ clusters. The number of clusters $k$ is a hyper-parameter and needs to be specified in advance. It is an unsupervised learning algorithm where all the instances are divided into clusters based on their distances from the cluster centers. The points in cluster $k_i$ are those whose distance from the cluster center $\mu_i$ is minimum among all the cluster centers. The algorithm terminates when the newly updated cluster centers are close enough (or equal) to the original cluster centers (before updation). After termination, each data point would have been assigned a label corresponding to the cluster in which they belong. Clustering is used for the defective instances to cluster them into $k$ clusters.

### 3.2.2 Over-sampling

Over-sampling involves the generation of new instances from the clusters which are formed. In this step, new instances are generated to balance the instances from each class. Let $N^0$ denote the total number of non-defective instances and $N^1$ denote the total number of defective instances. Therefore, the total number of new defective instances that need to be generated is $N = N^0 - N^1$. The defective instances have been divided into k clusters and let the number of defective instances in the $i^{th}$ cluster is $n_i$.
The new defective instances are generated between pairs of clusters. For $k$ clusters, there are $\frac{k \times (k-1)}{2}$ combinations.

For one combination of *p* and *q* cluster, we randomly sample one defective instance *i* from *p* cluster, and sample *j* from *q* cluster. A new defective instance *r* is determined by the interpolating *i* and *j*: $r = \delta \times i + \gamma \times j$. Where, $\delta = \frac{np}{np+nq}$ and $\gamma = \frac{nq}{np+nq}$. The number of new defective instances from this combination is calculated as $\frac{np+nq}{(k-1) \times N1} \times N$. This results in a diverse spread of the newly generated instances.

### 3.2.3 Noise Filtering

This method uses CLNI to remove the noise instances i.e instances that are near the decision boundary of the defective and non-defective instances. This is applied to the entire dataset and therefore both defective and non-defective instances are cleaned by this method.

CLNI method is based on the nature of the nearest neighbors of an instance. If the instance is a defective one and the majority of its neighbors are non-defective than the instance is a noisy one. The same holds for a non-defective instance as well. Therefore K-Nearest Neighbors, a supervised classification algorithm is used to identifying the neighborhoods for all the instances. The number of instances to be taken in the neighborhood *kn* is a hyperparameter and it needs to be specified in advance. The training dataset is thus constructed by the steps of clustering, over-sampling, and filtering. For computing results, we use Logistic Regression, Support Vector Machine, Random Forest, Naïve Bayes, and OS-ELM to evaluate the results.

## 3.3 Dataset

The analysis is carried out on the datasets PC3, PC4, and KC1 projects carried out by NASA. The original datasets are present at NASA Metrics Data Programme (MDP) website. They need to be cleaned to use them for analysis. The dataset used for our analysis is the cleaned versions available at https://github.com/klainfo/NASADefectDataset. The pre-processing steps carried out are those mentioned in [9].

## 4 EXPERIMENT
### 4.1 Design

The experiment comprises using 4 datasets PC4, PC3, PC1, and KC1, which are different projects carried out by the NASA group. Principal Component Analysis is used to extract features. The number of components used capture around 90% of the variance. A comparison is made on the results obtained on the normal dataset and the over-sampled dataset. For the original dataset, stratified 5-Fold cross-validation is used. For over-sampling 10 Folds are generated, and oversampling is done for the 9 out of 10 splits for all the parameters. Results are computed for the 10[th] Fold from all the parameters and averaged. For calculating results, Naïve Bayes Classifier, Logistic Regression, Support Vector Machines, Random Forest, and Online Sequential Learning Machine are used. The results are carried out on the default parameters for the implementation in the sklearn library and exceptions are specified in the following section. The parameters used for KMFOS algorithm are as follows:

- *k:* 3,5,20,50
- *kn:* 5,15,20

Average results are reported for each of the parameter settings along with the standard deviations.

### 4.2 Evaluation Metrics

Generally, one of the metrics used for classification is the accuracy score but when the dataset is imbalanced, the accuracy score can be misleading as naively predicting the majority class can yield a decent accuracy score. The measures used for analysis are *Recall Score* ad *Balanced accuracy score*.

*The recall* is the ratio of the number of correctly predicted defective instances to the total number of defective instances in the dataset. The bigger the recall value is, the better the performance of the method. It is defined as

$$Recall = \frac{TP}{TP+FN}$$

Where *TP* is True Positive, and *FN* is False Negative.
Balanced accuracy is defined as;

$$Balanced\ Accuracy = \frac{Acc0+Acc1}{2}$$

Where Acc0 denotes the accuracy of prediction for non-defective instances and Acc1 denotes the accuracy of prediction for defective instances. Higher the balance accuracy, the better would be the performance of the method.

## 5 RESULTS
### 5.1 Hyperparameters for Algorithms

- Logistic Regression
    - Solver = lbfgs
- SVM
    - Kernel = RBF
    - Gamma = auto
- Random Forest
    - N_estimators = 100
- OS-ELM
    - Neurons in the hidden layer: Approx 1.5 to 2 times the No. of input features
    - Activation: sigmoid
    - Dataset used for initial training = 1.5 times the number of neurons in the hidden layer.

**Inferences:**
- The average performance of SVM, Logistic Regression, and OS-ELM are similar however, the standard deviation for OS-ELM is the least among

these algorithms. This can be attributed to the hidden layer present in the OS-ELM which results in the algorithm learning complex decision boundaries. The average performance of OS-ELM is poor for the original imbalanced datasets which might be due to the algorithm underfitting for PC3, PC4 and overfitting for KC1 dataset[Fig.4(a)-(e) ]

- The performance of Naïve Bayes (Gaussian) and Random Forest were lesser as compared to SVM, Logistic Regression, and OS-ELM. As discussed earlier, there are many hyperparameters for Random Forest which need to be tuned for getting optimal results but are out of the scope of this report. Naïve Bayes (Gaussian), ignores any considers different features to be independent of each other while predicting the value of the target variable however, this isn't the case with software-defect data. The features exhibit significant correlation and therefore, this can be leveraged by a learning algorithm.

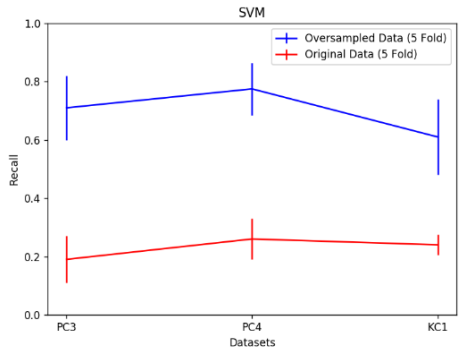

**Fig.4a:** SVM

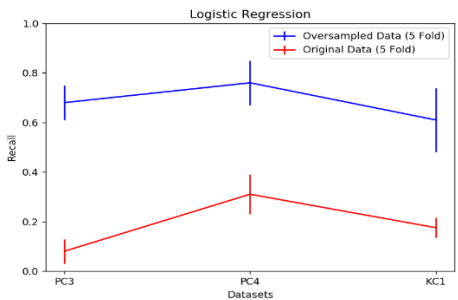

**Fig.4b:** LR

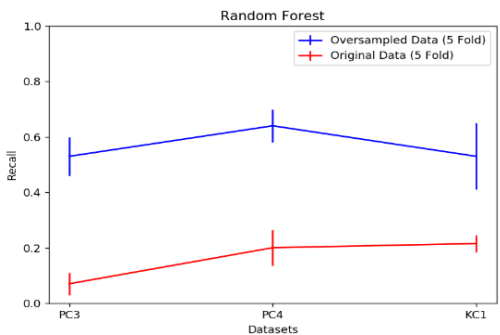

**Fig.4c:** RF

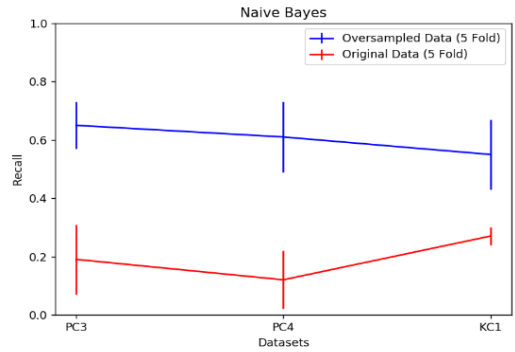

**Fig.4d:** NB

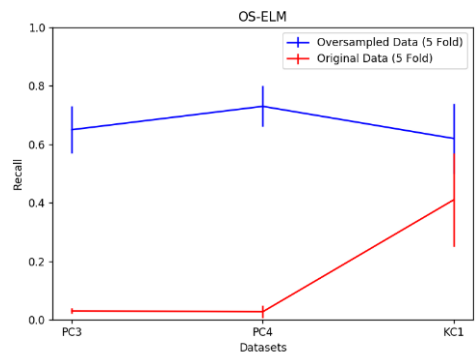

**Fig.4e:** OSELM

### 5.3 PC4 Dataset

The results obtained on the PC dataset before and after using KFMOS. The specification of the dataset is:

- Number of Instances – 16962
- Defective Instances – 502
- % of Instances Defective – 2.96%

**Table-1**
**Performance Assessments against PC4 Dataset**

| Classifier | Metric | Original | Over-sampled |
|---|---|---|---|
| SVM | Recall | $\mu = 0.259$ $\sigma = 0.072$ | $\mu = 0.775$ $\sigma = 0.093$ |
| | Balanced accuracy | $\mu = 0.616$ $\sigma = 0.036$ | $\mu = 0.762$ $\sigma = 0.075$ |
| Logistic Regression | Recall | $\mu = 0.315$ $\sigma = 0.081$ | $\mu = 0.764$ $\sigma = 0.090$ |
| | Balanced Accuracy | $\mu = 0.641$ $\sigma = 0.044$ | $\mu = 0.765$ $\sigma = 0.072$ |
| Random Forest | Recall | $\mu = 0.202$ $\sigma = 0.065$ | $\mu = 0.641$ $\sigma = 0.063$ |
| | Balanced Accuracy | $\mu = 0.582$ $\sigma = 0.029$ | $\mu = 0.736$ $\sigma = 0.060$ |
| Naïve Bayes | Recall | $\mu = 0.117$ $\sigma = 0.107$ | $\mu = 0.610$ $\sigma = 0.118$ |
| | Balanced | $\mu = 0.534$ | $\mu = 0.714$ |

| | Accuracy | σ = 0.043 | σ = 0.048 |
|---|---|---|---|
| OS-ELM | Recall | μ = 0.095 σ = 0.021 | μ = 0.731 σ = 0.066 |
| | Balanced Accuracy | μ = 0.543 σ = 0.011 | μ = 0.737 σ = 0.058 |

### 5.3 PC3 Dataset

The results obtained on the PC4 dataset before and after using KFMOS. The specification of the dataset is:

- Number of Instances – 1379
- Defective Instances – 178
- % of Instances Defective – 12.9%

**Table-2**
**Performance Assessments against PC3 Dataset**

| Classifier | Metric | Original | Over-sampled |
|---|---|---|---|
| SVM | Recall | μ = 0.189 σ = 0.087 | μ = 0.721 σ = 0.155 |
| | Balanced accuracy | μ = 0.550 σ = 0.051 | μ = 0.709 σ = 0.077 |
| Logistic Regression | Recall | μ = 0.080 σ = 0.053 | μ = 0.678 σ = 0.072 |
| | Balanced Accuracy | μ = 0.527 σ = 0.033 | μ = 0.704 σ = 0.057 |
| Random Forest | Recall | μ = 0.073 σ = 0.040 | μ = 0.529 σ = 0.067 |
| | Balanced Accuracy | μ = 0.519 σ = 0.023 | μ = 0.664 σ = 0.030 |
| Naïve Bayes | Recall | μ = 0.189 σ = 0.120 | μ = 0.656 σ = 0.084 |
| | Balanced Accuracy | μ = 0.562 σ = 0.064 | μ = 0.666 σ = 0.048 |
| OS-ELM | Recall | μ = 0.027 σ = 0.012 | μ = 0.651 σ = 0.076 |
| | Balanced Accuracy | μ = 0.509 σ = 0.004 | μ = 0.695 σ = 0.039 |

### 5.5 KC1 Dataset

The results obtained on the KC1 dataset before and after using KFMOS. The specification of the dataset is:
- Number of Instances – 1379
- Defective Instances – 178
- % of Instances Defective – 12.9%

**Table-3**
**Performance Assessments against KC1 Dataset**

| Classifier | Metric | Original | Over-sampled |
|---|---|---|---|
| SVM | Recall | μ = 0.237 σ = 0.035 | μ = 0.608 σ = 0.132 |
| | Balanced accuracy | μ = 0.555 σ = 0.024 | μ = 0.694 σ = 0.022 |
| Logistic Regression | Recall | μ = 0.175 σ = 0.037 | μ = 0.613 σ = 0.130 |
| | Balanced Accuracy | μ = 0.575 σ = 0.017 | μ = 0.700 σ = 0.021 |
| Random Forest | Recall | μ = 0.215 σ = 0.032 | μ = 0.535 σ = 0.040 |
| | Balanced Accuracy | μ = 0.579 σ = 0.011 | μ = 0.662 σ = 0.045 |
| Naïve Bayes | Recall | μ = 0.274 σ = 0.033 | μ = 0.550 σ = 0.120 |
| | Balanced Accuracy | μ = 0.607 σ = 0.025 | μ = 0.684 σ = 0.036 |
| OS-ELM | Recall | μ = 0.411 σ = 0.090 | μ = 0.621 σ = 0.123 |
| | Balanced Accuracy | μ = 0.519 σ = 0.041 | μ = 0.694 σ = 0.025 |

## 6 THREATS TO VALIDITY: INTERNAL & EXTERNAL

The use of learning algorithms didn't involve hyper-parameter tuning, thus consequently, the results obtained may be biased. The metrics used for the project are *recall* and *balanced accuracy* however, other metrics such as *G-means, MCC,* and *F1-score* aren't used. The projects used for analysis are widely used in software defect prediction research but there are questions regarding their generalizability to commercial software projects.

## 7 CONCLUSION

It can be concluded that using KMFOS oversampling technique significantly enhances the performance of the classifier which is to be used. KMFOS ensures that the generated instances are diverse enough and solve the boundary ambiguity problem. The researchers found this method to be superior to many state-of-the-art sampling-based techniques such as SMOTE, ADASYN, Borderline – SMOTE, etc.

P.Khakar and, R.K Dubey et al.: Integrity of Machine Learning Algorithms against Software Defect Prediction

7878_887.

[5] H. Li, P. Zou, X. Wang, and R. Z. Xia, "*A new combination sampling method for imbalanced data*" in Proc. Chin, Intell. Automat. Conf., in Lecture Notes in Electrical Engineering, vol. 256, 2013, pp. 547_554.

[6] H. He, Y. Bai, E. A. Garcia, and S. Li, "*ADASYN: Adaptive synthetic sampling approach for imbalanced learning*" in Proc. IEEE Int. Joint Conf. Neural Netw. (IEEE World Congr. Comput. Intell. Hong Kong, Jun 2008, pp. 1322_1328.

[7] Zhang, J. & Mani, I. (2003). KNN Approach to Unbalanced Data Distributions: A Case Study Involving Information Extraction. *Proceedings of the ICML'2003 Workshop on Learning from Imbalanced Datasets*,

[8] K. Gowda and G. Krishna, "The condensed nearest neighbor rule using the concept of the mutual nearest neighborhood (Corresp.)," in *IEEE Transactions on Information Theory*, vol. 25, no. 4, pp. 488-490, July 1979, DOI: 10.1109/TIT.1979.1056066

[9] M. Shepperd, Q. Song, Z. Sun and C. Mair, "Data Quality: Some Comments on the NASA Software Defect Datasets," in *IEEE Transactions on Software Engineering*, vol. 39, no. 9, pp. 1208-1215, Sept. 2013, DOI: 10.1109/TSE.2013.11.